\documentclass{article}

\usepackage[english]{babel}

\usepackage[letterpaper,top=2cm,bottom=2cm,left=3cm,right=3cm,marginparwidth=1.75cm]{geometry}

\usepackage{amsmath}
\usepackage{amssymb}
\usepackage{caption}
\usepackage{subcaption}
\usepackage[overload]{empheq}
\usepackage{float}
\usepackage{graphicx}
\usepackage[colorlinks=true, allcolors=blue]{hyperref}
\usepackage{multirow}
\usepackage[section]{placeins}
\usepackage{textcomp} 

\title{NET-TEN: a silicon neuromorphic network for low-latency detection of seizures in local field potentials}
\author{Margherita Ronchini, Yasser Rezaeiyan, Milad Zamani,\\Gabriella Panuccio, Farshad Moradi}

\begin{document}
\maketitle

\begin{abstract}
Therapeutic intervention in neurological disorders still relies heavily on pharmacological solutions, while the treatment of patients with drug resistance remains an open challenge. This is particularly true for patients with epilepsy, 30\% of whom are refractory to medications. Implantable devices for chronic recording and electrical modulation of brain activity have proved a viable alternative in such cases. To operate, the device should detect the relevant electrographic biomarkers from Local Field Potentials (LFPs) and determine the right time for stimulation. To enable timely interventions, the ideal device should attain biomarker detection with low latency while operating under low power consumption to prolong the battery life. Neuromorphic networks have progressively gained reputation as low-latency low-power computing systems, which makes them a promising candidate as processing core of next-generation implantable neural interfaces. Here we introduce a fully-analog neuromorphic device implemented in CMOS technology for analyzing LFP signals in an \textit{in vitro} model of acute ictogenesis. We show that the system can detect ictal and interictal events with ms-latency and with high precision, consuming on average 3.50~nW during the task. Our work paves the way to a new generation of brain implantable devices for personalized closed-loop stimulation for epilepsy treatment.
\end{abstract}

\section{Introduction}

Pharmaceutical treatment currently represents the prevailing therapy in epilepsy. Yet, about one-third of the patients fail to respond to anti-epileptic drugs \cite{kwan2000early,chen2018treatment}. Resorting to other strategies is therefore necessary in these cases to prevent or suppress seizures. Deep Brain Stimulation (DBS) holds great promise for treating medically intractable epilepsy \cite{fisher2010electrical,morrell2011responsive,jobst2017brain,li2018deep,elder2019responsive}. In this regard, a distinction must be made between open-loop and closed-loop DBS. In the former approach, the stimulation is either continuous or cyclic; the electrical stimulus is repeated periodically following a pre-programmed schedule, regardless of the dynamic state of the targeted brain circuitry. Conversely, closed-loop devices operate adaptively, delivering a stimulation pattern upon the detection of specific electrographic biomarkers. Both methods can effectively reduce the duration and/or the frequency of seizure \cite{fisher2010electrical,morrell2011responsive}. However, empirical evidence points to a higher efficacy and fewer adverse effects of the closed-loop over the open-loop paradigm \cite{good2009control,salam2015seizure}, as the stimulation is informed by the ongoing brain activity \cite{muller2017neurotechnology}.\par
A closed-loop system requires the integration of three essential components: (1) a recording interface to amplify and filter the signal, (2) a processing unit to analyze the recorded signals and extract informative features, and (3) a stimulation back-end to deliver an electrical feedback. The processor plays a pivotal role, since success in anticipating and averting seizures depends on its responsiveness and accuracy. Highly accurate tracking of seizures can be achieved offline using software-based algorithms \cite{tzallas2009epileptic,ghosh2009new}, but their computational complexity forces them to run offline on high-performance computers \cite{salam2011novel}. Local real-time analysis of data is therefore necessary to reduce wireless communication power overhead  \cite{verma2010micro}, though on-chip processing tightens the already stringent requirements in terms of area and power consumption \cite{altaf20151}. In this respect, plenty of wearable and implantable classification systems have been proposed \cite{verma2010micro,salam2011novel,chen2013fully,mirzaei2013fully,altaf20151}. Unfortunately, most devices still lack an intelligent control algorithm able to work around the large inter-patient variability \cite{parastarfeizabadi2017advances}. To address all these challenges, neuromorphic  processing cores promise to become an integral part of next-generation neural implants.\par
Neuromorphic circuits operate on the same principles as biological information-processing systems, drawing inspiration from the physical phenomena that govern the electrical behavior of neurons and synapses to implement computational primitives \cite{mead1990neuromorphic}. Thanks to the spike-based representation of information, the parallelism of multiple processing elements and their colocalization with memory units, neuromorphic systems achieve low-latency and low-power performance, for which they have emerged as a worthy opponent for von Neumann architectures in computing systems \cite{indiveri2015memory}. These features also make them an appealing candidate for implantable neural interfaces, as their structure is inherently suited to simulating spiking neural networks, which are a powerful tool to solve problems of spatiotemporal pattern recognition \cite{kasabov2013dynamic,moraitis2020short,she2021heterogeneous,debat2021event}. As such, neuromorphic systems have already found successful application for processing a broad range of biological electrical signals \cite{corradi2015neuromorphic,bauer2019real,corradi2019ecg,donati2019discrimination,ceolini2020hand,ma2020emg}, including high-frequency oscillations as pathological biomarkers of epilepsy \cite{ burelo2021spiking,sharifshazileh2021electronic,burelo2022neuromorphic} and seizure detection \cite{ ronchini2021cmos}.\par 
Here, we describe NET-TEN, a fully-analog subthreshold neuromorphic network implemented in a standard 180 nm Complementary Metal-Oxide-Semiconductor (CMOS) process for the detection of ictal and interictal events from rodent brain slices coupled to Multi-Electrode Array (MEA).

\section{Results}

\subsection{NET-TEN implementation}

\begin{figure}[ht!]
\centering
    \includegraphics[width=\textwidth,trim={3.8cm 1.5cm 2.4cm 0},clip]{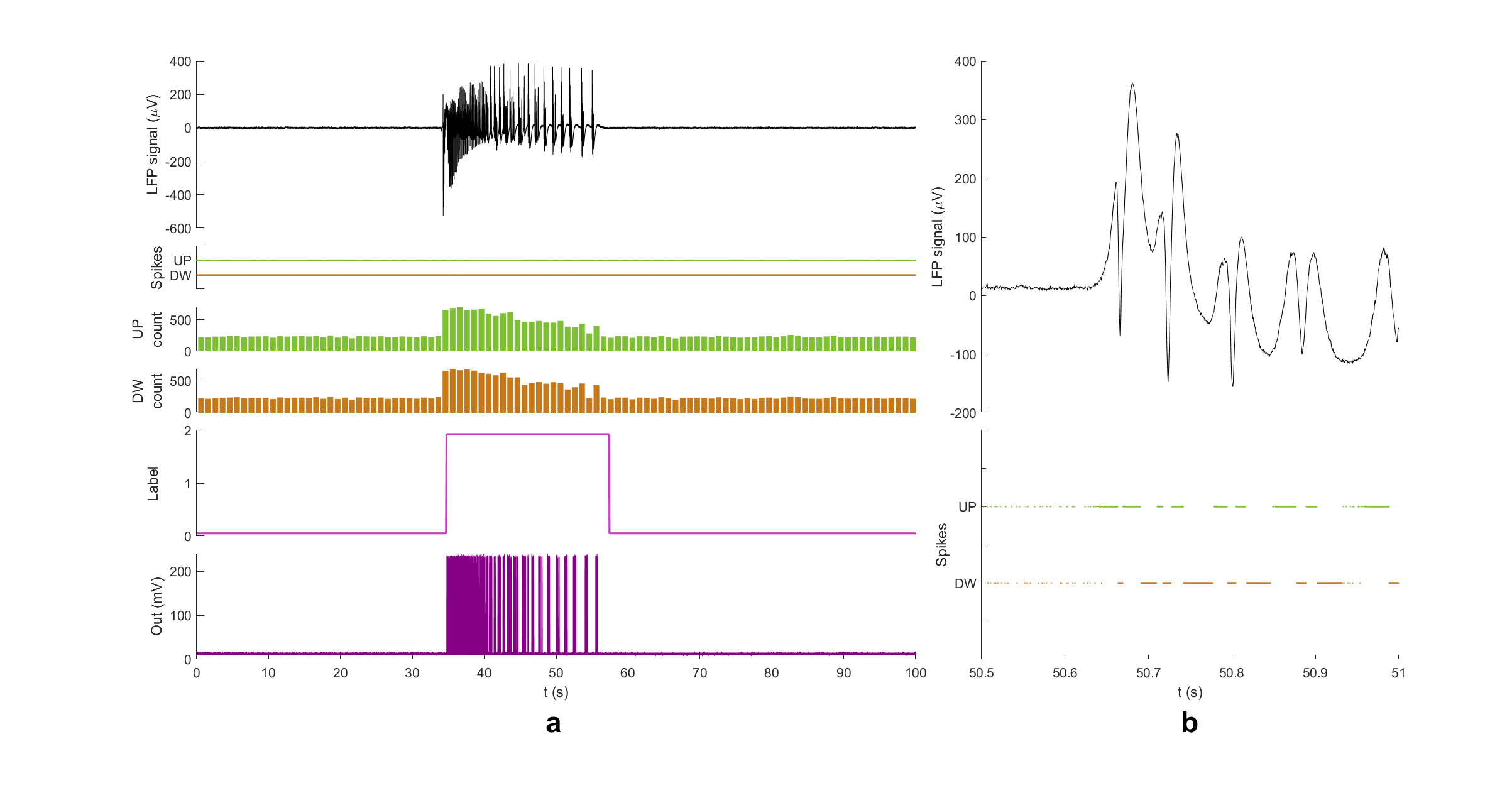}
    \caption{Example of a detection task completed by NET-TEN. a) From top to bottom: A 100 seconds long sample of prerecorded Local Field Potential (LFP) data. One single ictal event emerges from the baseline; UP and DW spikes generated from the conversion of the LFP data by the Step-Forward Encoding (SFE) algorithm and used as input to NET-TEN; Graphical representation of the density of UP and DW spikes fed to NET-TEN. The height of each bar corresponds to the number of spikes accumulated within one second; Label as manually assigned (0 = baseline, 1 = interictal, 2 = ictal); Voltage signal of one of the output neurons of NET-TEN as measured by the oscilloscope. b) Zoom-in of the LFP signal with the corresponding UP and DW spikes generated by the SFE algorithm. Positive slopes in the signal lead to the emission of UP spikes; negative slopes result in DW spikes.}
    \label{fig:Ictal}
\end{figure}

Figure \ref{fig:Ictal} shows the results of a detection task performed by NET-TEN. Prerecorded Local Field Potential (LFP) data of 100-second duration were converted into spikes in software using the Step-Forward Encoding (SFE) algorithm proposed by Kasabov et al. \cite{kasabov2016evolving} and delivered to the fabricated network through an arbitrary waveform generator. The encoding process gives rise to two spike trains: UP spikes, associated with positive-going signal deflections, and DW spikes, associated with negative-going signal deflections. A second waveform generator synchronized with the first one was employed to carry the label signal and be able to display it on the oscilloscope together with NET-TEN neuronal spiking. Firing activity at the output signaled the detection of a pathological pattern in the LFP recordings.\par

\begin{figure}[ht!]
\centering
    \includegraphics[width=\textwidth]{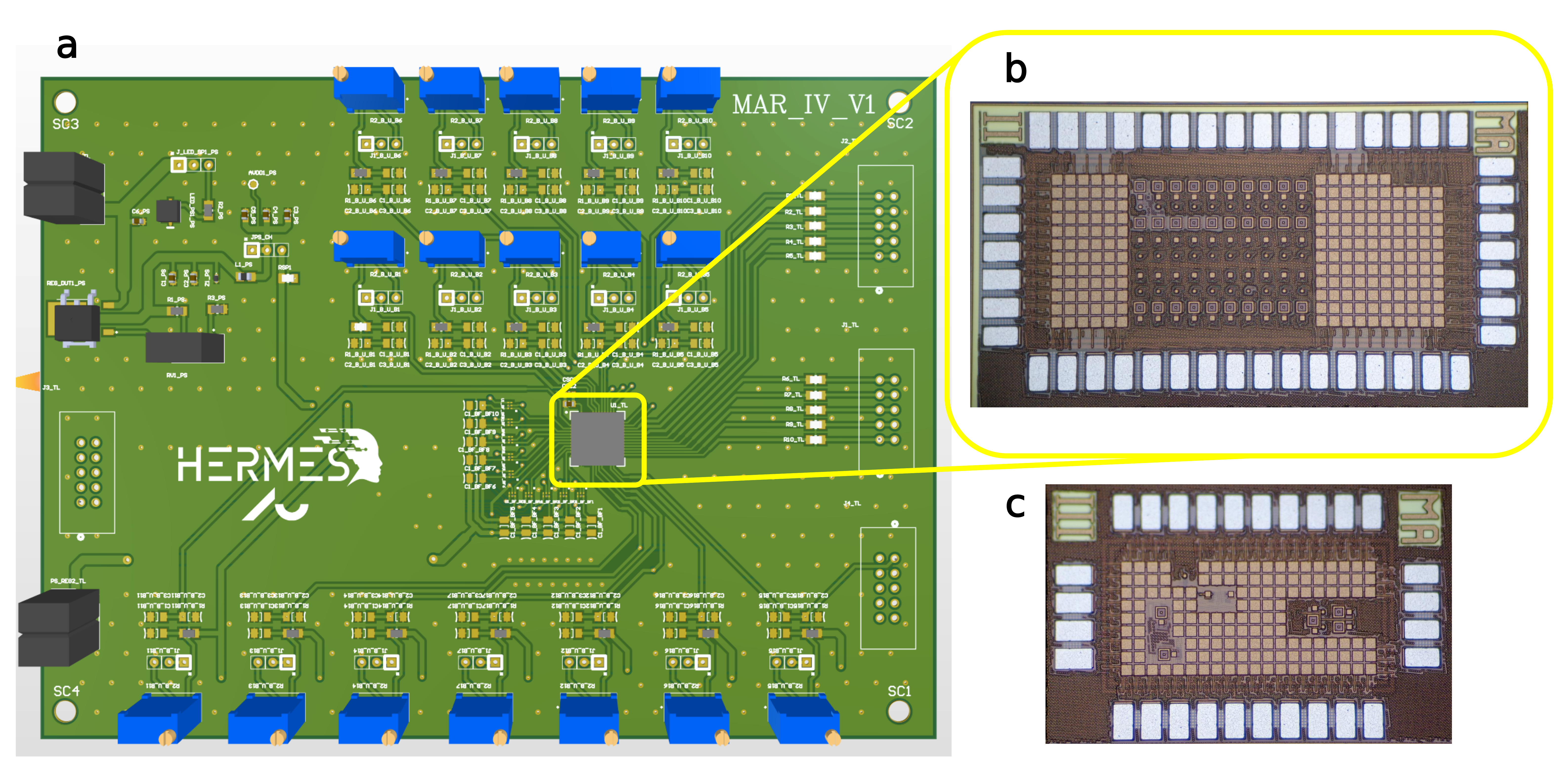}
        \caption{a) Rendering of the prototyping Printed Circuit Board (PCB) comprehensive of potentiometers to adjust the circuit biases. The 48-pin Quad Flat No-Lead (QFN48) footprint outlined in yellow hosts the packaged NET-TEN chip. b) Photograph of the fabricated NET-TEN circuit die. c) Die containing the stand-alone components: neuron block and synapse.}
    \label{fig:PCB_Layout}
\end{figure}

Figure \ref{fig:PCB_Layout} provides a summary of the designed hardware components. The neuromorphic system was implemented in a 180nm CMOS technology node and its layout covered an area of 1.08~$mm^2$. The circuit voltage biases were adjusted by tuning the relative potentiometers on the Printed Circuit Board (PCB, depicted in Figure \ref{fig:PCB_Layout}a), while at the same time the corresponding firing activity generated at the output was compared with the label. In this way, it was possible to determine heuristically in which direction to steer each bias, in order to improve the classification accuracy. The fabricated chip is portrayed in Figure \ref{fig:PCB_Layout}b. The voltage supply was set at 250~mV. The total static power consumption was 0.68~pW. The average power consumed during a detection task calculated across samples was 3.50~nW. To evaluate the electrical behavior of the various functional modules that make up the neuromorphic network, these were fabricated as stand-alone components on a separate chip (Figure \ref{fig:PCB_Layout}c).\par

\begin{figure}[ht!]
\centering
    \includegraphics[width=\textwidth]{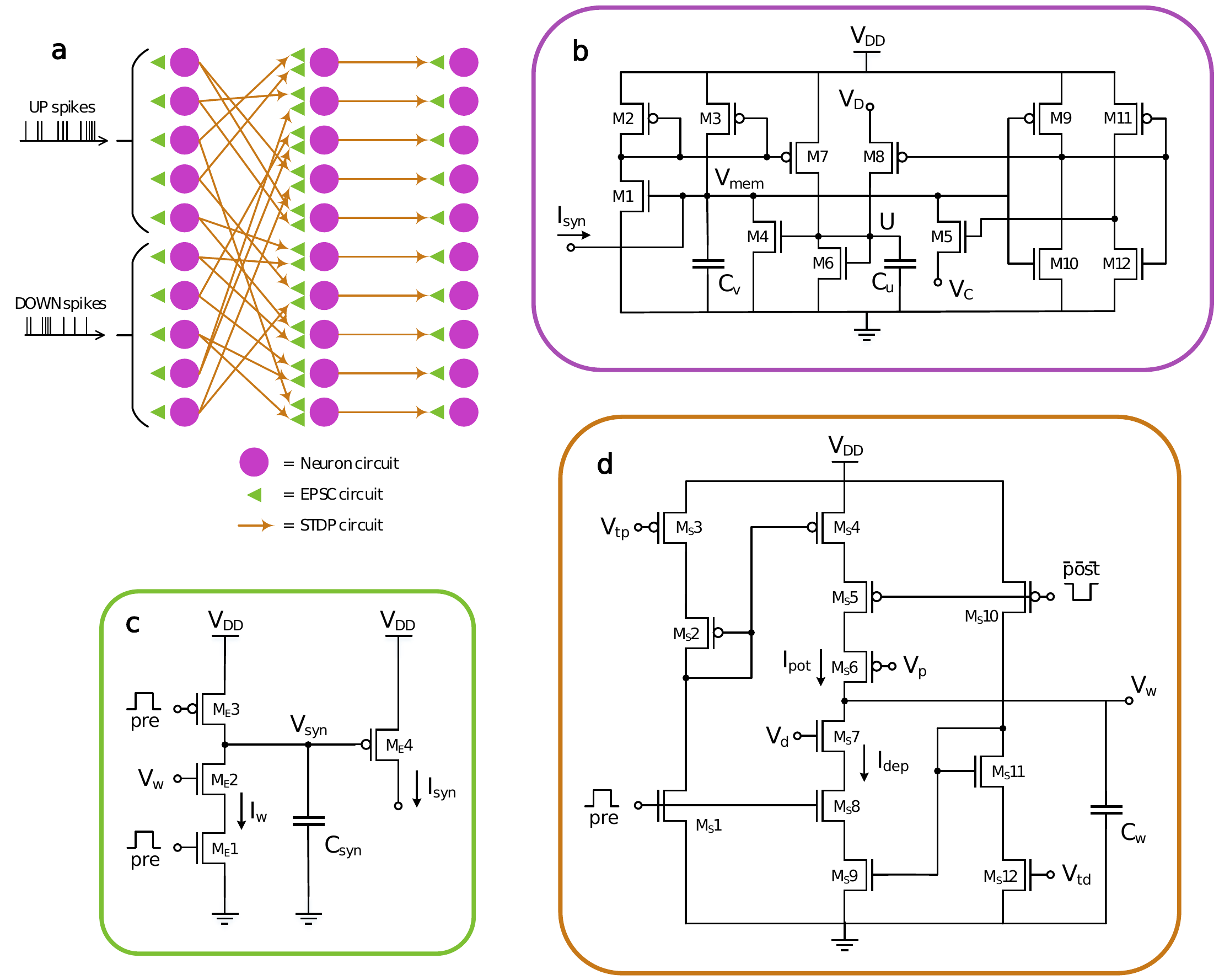}
    \caption{Schematic view of NET-TEN building blocks. a) Neuron circuit. It can be functionally broken down into three sub-circuits: one controlling the transmembrane voltage $V_{mem}$, another one in charge of the slow recovery variable $U$ and finally a comparator (composed by two inverters), that compares $V_{mem}$ to a threshold and generates an electric pulse (spike) if this is exceeded. $V_{C}$ and $V_{D}$ are the after-spike reset parameters, which are used to obtain the various possible firing behaviors of cortical neurons. b) Excitatory Post-Synaptic Current (EPSC) circuit produces the current $I_{syn}$ that stimulates the neuron. The current $I_{syn}$ has a rapid onset and then slowly decays with time, and its peak is a function of the synaptic weight $V_{w}$. c) Spike-Timing Dependent Plasticity (STDP) circuit, proposed by Indiveri \cite{indiveri2003neuromorphic} and here implemented fully in sub-threshold. d) NET-TEN architecture. The network is composed by three layers of ten neurons each. There are no STDP blocks in correspondence of the input layer and the current injected into each input neuron is gated by a static weight set from the outside, on the PCB. Every neuron of the hidden layer receives two connections from the previous layer. Each connection corresponds to a STDP circuit with its respective EPSC circuit. The currents delivered by the two separate EPSC blocks (one EPSC per connection) are summed together and integrated onto the capacitor $C_{v}$ of the hidden neuron.}
    \label{fig:Schematic}
\end{figure}

Figure \ref{fig:Schematic} illustrates the schematic of the three principal building blocks that constitute NET-TEN, namely the neuron, the Excitatory Post-Synaptic Current (EPSC) and the Spike-Timing Dependent Plasticity (STDP) circuits. As can be seen in Figure \ref{fig:Schematic}a, NET-TEN network has a feed-forward architecture composed by three sparsely-connected layers of ten neurons each. Every neuron of the input layer project to two neurons of the hidden layer, whose neurons also have a fan-in of two. Hidden and output layers are connected together in a one-to-one fashion, meaning every neuron of the hidden layer forms a synapse with one and only one neuron of the output. \par 

The neuron, depicted in Figure \ref{fig:Schematic}b, was initially proposed in \cite{ronchini2020tunable} and further described in \cite{ronchini2021cmos}. The circuit replicates biologically plausible dynamics, following the Izhikevich model. After performing the Monte Carlo analysis, the transistor sizes were modified with respect to the original implementation and the capacitance values were set to $C_{v}$ = 76.86~fF and $C_{u}$ = 614.88~fF, to alleviate the effect of process variations. One single neuron cell occupied an area of 1476.84~$\mu{m}^2$. In the schematic, the two state variables of the Izhikevich model, i.e. the transmembrane voltage $V_{mem}$ and the slow recovery variable $U$, are represented by the voltage across the integrating capacitors $C_{v}$ and $C_{u}$, respectively. $I_{syn}$ is the synaptic current generated on the basis of the previous network layer and, when injected into the neuron, it perturbs its equilibrium. $M1-M3$ current mirror serves as a positive feedback to accelerate the rise of $V_{mem}$ and bring it closer to the switching threshold of $M9-M10$ inverter. This increase is counteracted by the leakage current $I_{M4}$, proportional to $U$, which causes the probability of firing to be at a minimum immediately after a spike. Another current mirror, formed by the transistors $M1, M2$ and $M7$, couples $U$ to $V_{mem}$. A negative feedback introduced by transistor $M6$ discharges $C_{u}$, reducing $U$ by an amount dependent on its own present value. The two inverters $M9-M10$ and $M11-M12$ function as a comparator with a fixed threshold, whose value derives from how the transistors are sized. When the transmembrane voltage exceeds the threshold, the comparator emits a negative pulse to turn on $M8$ and reset $U$, and a positive pulse to reset $V_{mem}$ through $M5$. The reset mechanism is controlled by voltage biases $V_{C}$ and $V_{D}$, which are a means to alter the spiking pattern. Negative and positive pulses are harnessed at the network level to activate the synaptic circuits whenever a pre- or post-synaptic spike is elicited  \cite{ronchini2021cmos}.\par

The EPSC circuit shown in Figure \ref{fig:Schematic}c is responsible for inducing the post-synaptic current $I_{syn}$ when triggered by a pre-synaptic spike and it constitutes the fundamental module of the silicon synapse. The EPSC is a low-pass filtered version of the original spike impulse, and presents a steep rising phase and an approximately exponential decay. In between spikes, the transistor $M_{E}3$ is on and conducts a current. Since at the same time $M_{E}1$ is off, no current can flow in this branch and therefore all $I_{M_{E}3}$ is diverted into $C_{syn}$, precharging the node $V_{syn}$ to $V_{DD}$. When a pre-synaptic pulse arrives, the discharge phase starts. As $M_{E}1$ turns on and $M_{E}3$ deactivates, a current $I_{w}$ is drawn from $C_{syn}$. The amplitude of such current is directly proportional to the synaptic strength $V_{w}$, so the greater is $V_{w}$ the more $C_{syn}$ is discharged. At the falling edge of the pulse $pre$, $M_{E}3$ begins to conduct current again, leading to a slow increase in $V_{syn}$ value. The whole circuit operates in subthreshold, which means that the EPSC has an exponential dependence on $V_{syn}$. The ratio between the sizing of $M_{E}1$ and that of $M_{E}3$ establishes the time course of $I_{syn}$, resulting in a fast onset followed by a slow decaying phase (corresponding to the recharge of $C_{syn}$ after spike completion). The EPSC circuit response has been derived analytically in \cite{ronchini2021cmos}, which also illustrates the relationship between $I_{syn}$ peak value and the synaptic weight $V_{w}$. Simulations of the circuit with a power supply $V_{DD} = 250$~mV, meaning the same one used in the experimental studies, and a static weight $V_{w} = 200$~mV, which is the maximum bias value achievable with the tunable voltage dividers present on the PCB, led to an EPSC with peak current $I_{syn_{peak}} = 486.39$~pA and a decay constant of $\tau = 212.83$~$\mu s$. The layout of the EPSC was 462.09~$\mu{m}^2$ large.\par 
The synaptic strength for each of the connections that link NET-TEN layers together is dynamically updated in agreement with the STDP rule and it can take up any analog value between $V_{DD}$ and ground. The STDP is an unsupervised learning algorithm modeled after biological evidence \cite{markram1997regulation}. NET-TEN featured the STDP CMOS implementation conceived by Indiveri \cite{indiveri2003neuromorphic}, which was here redesigned to work at ultra-low power (Figure \ref{fig:Schematic}d). Amplitude and temporal evolution of potentiation and depression could be regulated respectively by voltage biases $V_{tp}$ and $V_{td}$, and $V_{pb}$ and $V_{db}$. The new implementation occupied 1054.75~$\mu{m}^2$. Given that the synaptic weight was stored on a capacitor $C_{w}$ = 206.30~fF, the synapses were able to retain the computed value only for a short period of time, in the order of tens of milliseconds.

\subsection{Detection performance} 

\begin{figure}[h]
    \centering
    \includegraphics[width=\textwidth]{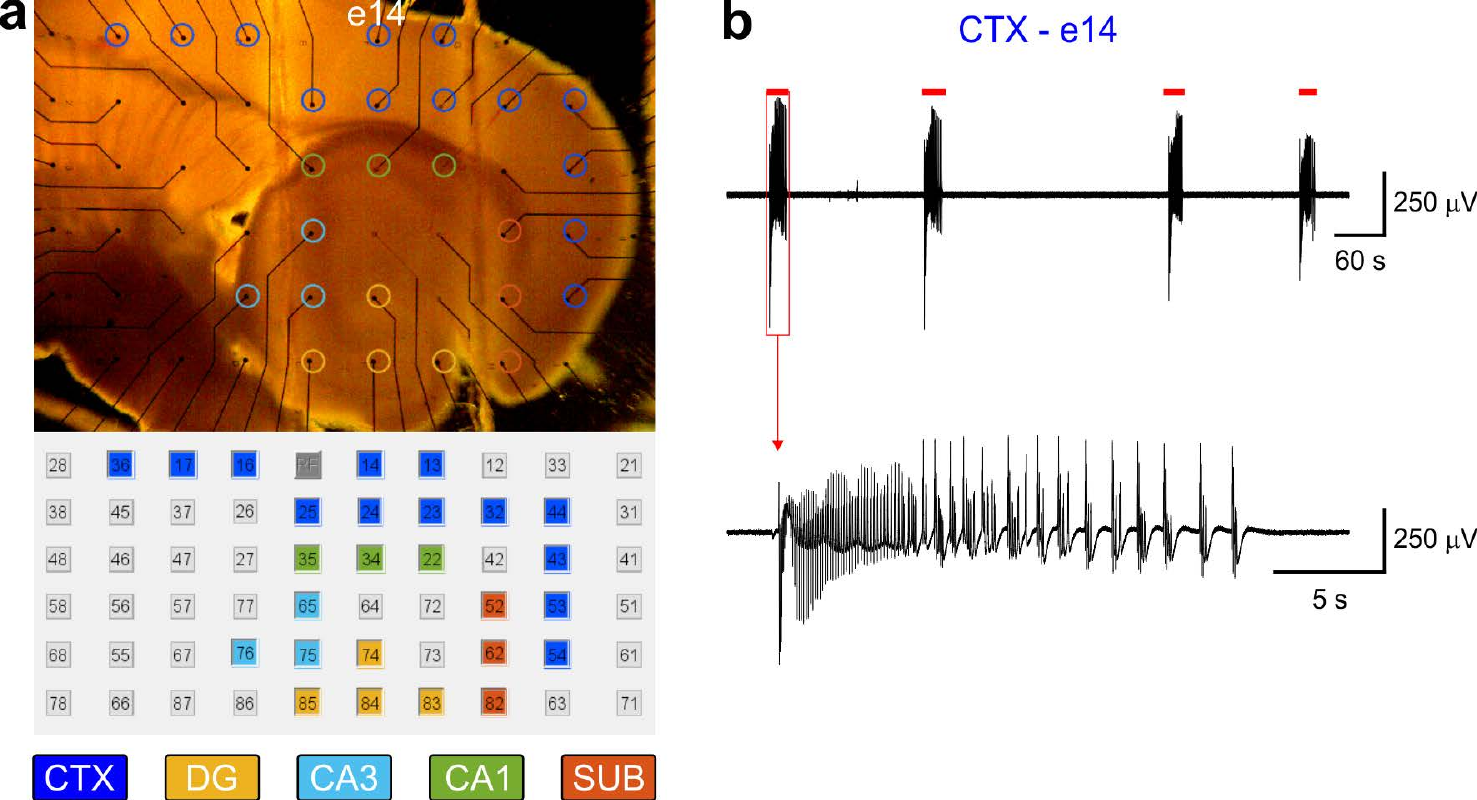}
    \caption{Multi-Electrode Array (MEA) electrode mapping. a) Snippet of the MATLAB GUI used to map the electrodes in contact with the regions of interest within the brain slice. On top is the picture of a mouse hippocampus-CTX slice coupled to a 6x10 MEA, where the mapped electrodes and the corresponding brain regions are identified by color-coded circles. Below the brain slice picture, a pushbutton panel enables the selection of the electrodes of interest (same color-coding as in the brain slice picture). b) Representative signal recorded from the brain slice in (a), electrode 14 (e14), showing recurrent ictal discharges (marked by the red bars). Below is the ictal discharge marked by the red box, visualized at expanded time scale to show the tonic-clonic-like features of ictal activity. CTX: parahippocampal cortex. DG: Dentate Gyrus. CA3: Cornu Ammonis 3. CA1: Cornu Ammonis 1. SUB: Subiculum.}
    \label{fig:electrode}
\end{figure}

The analyzed dataset comprises MEA recordings from the parahippocampal cortex (CTX) of n = 10 brain slices from n = 10 mice, to ensure fair representation of biological variability. The extracted samples were 100 s long signal segments around an ictal event. Three samples also featured interictal discharges, while one of them only contained an artifact. The dataset content is summarized in Table~\ref{tab:delay}. Figure \ref{fig:electrode} depicts a representative hippocampus-CTX slice coupled to a MEA and representative of the epileptiform patterns recorded by the electrode marked by the blue circle. During the measurements, the UP spikes were fed to five input neurons, the DW spikes to the other five, as shown in Figure \ref{fig:Schematic}d. All ten output pins were probed with the oscilloscope, but only four out of these ten were deemed to be informative for pattern recognition. In fact, despite all neuron and synapse circuits share the same nominal biases values, device mismatch affects the individual instances behavior. The described results refer to said four output channels. \par

\begin{figure}[ht!]
\centering
    \includegraphics[width=\textwidth,trim={2.8cm 0 2.2cm 0},clip]{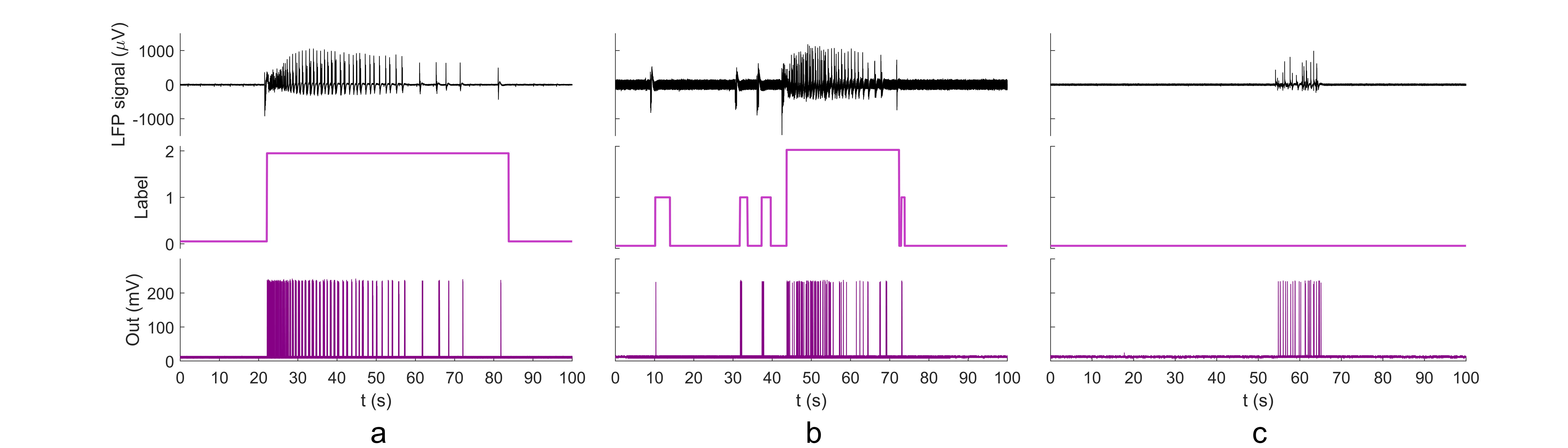}
    \caption{NET-TEN response in three different scenarios. The Local Field Potential (LFP) signals are presented, together with the corresponding label (0 = baseline, 1 = interictal, 2 = ictal) and the experimental measurement of one output of the fabricated neuromorphic device. All the displayed recordings belong to the same output neuron. a) One single ictal event. The spiking activity at the output is initially very dense and it gradually decreases with time, in correspondence of the bursting phase. b) Three interictal spikes building up to an ictal discharge. Another interictal spike follows the tail of the ictal episode. c) Artifact.}
    \label{fig:I_II_A}
\end{figure}

\begin{figure}[ht!]
\centering
    \includegraphics[width=\textwidth,trim={2.6cm 0.5cm 3.5cm 0},clip]{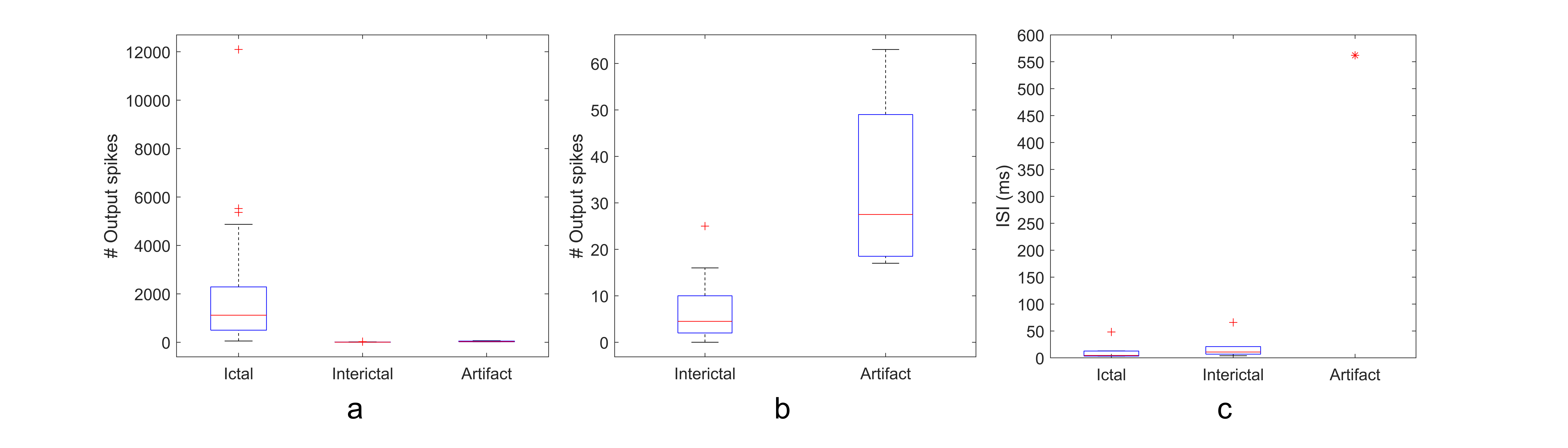}
    \caption{a) Boxplot of the number of spikes at the device output grouped by category. b) A side-by-side comparison between the number of output spikes for interictal discharge and artifact. c) Boxplot of the interspike interval (ISI) for the three categories. Only one observation is available for the artifact as the ISI value as been averaged across output channels.}
    \label{fig:Boxplot}
\end{figure}

NET-TEN was able to successfully recognize abnormal patterns, firing only in correspondence of pathological activity at the input. The output response exhibited by the network in three distinct cases can be observed in Figure \ref{fig:I_II_A}. The precision score was 0.998. The device detected ictal events with a mean delay of 110.4 ms, while it took about 126.6 ms to detect an interictal discharge. Table \ref{tab:delay} details the detection delays calculated for each LFP sample and for each retrieved output signal. Observe how the detection delay varied among the output channels, with some channels more reactive and others less reactive to the stimulus, depending on the internal dynamics of the network. Further classification capabilities were explored, looking for elements that would allow a distinction between ictal and interictal occurrence, and possible ways to discard artifacts. The amount of spikes released at the output differed between categories, and so did the interspike interval (ISI) in-between the first two spikes. Boxplots for both metrics are provided in Figure \ref{fig:Boxplot}. The median spike count for ictal events amounted to 1117.5 with an interquartile range (IQR) of 1762.5. For interictal activity, the median spike count was 4.5 with IQR 7. In the case of the artifact, median and IQR were 27.5 and 22.8, with the first quartile being 19.3 and the third quartile equal to 42. Coming to the ISI, the values were first averaged across output channels. The median ISI was 4.3~(3.7-4.8)~ms for ictal and 7.8~(3.8-33.4)~ms for interictal patterns. Instead, since there was only one sample including one artifact, the outcome of the average of ISI across measured output signals came down to a single value corresponding to $ISI_{artifact} = 562~ms$.

\begin{table}[ht!]
\caption{\label{tab:delay} Ictal and interictal detection delay for the different samples. *Multiple interictal discharges present in the sample. The one displayed is the average value of the delays of individual discharges.}
\resizebox{\textwidth}{!}{
\begin{tabular}{cc|cccc|cccc}
\\
\multicolumn{1}{c|}{\multirow{2}{*}{\textbf{Sample}}} & \multirow{2}{*}{\textbf{Type}} & \multicolumn{4}{c|}{\textbf{Ictal detection delay (ms)}} & \multicolumn{4}{c}{\textbf{Interictal detection delay (ms)}} \\ \cline{3-10} 
\multicolumn{1}{c|}{} &  & \multicolumn{1}{c|}{Ch6} & \multicolumn{1}{c|}{Ch7} & \multicolumn{1}{c|}{Ch8} & Ch10 & \multicolumn{1}{c|}{Ch6} & \multicolumn{1}{c|}{Ch7} & \multicolumn{1}{c|}{Ch8} & Ch10 \\ \hline
\multicolumn{1}{c|}{1} & Ictal & \multicolumn{1}{c|}{31.4} & \multicolumn{1}{c|}{31.2} & \multicolumn{1}{c|}{31.6} & 52.9 & \multicolumn{1}{c|}{-} & \multicolumn{1}{c|}{-} & \multicolumn{1}{c|}{-} & - \\
\multicolumn{1}{c|}{2} & Ictal + Interictal & \multicolumn{1}{c|}{48.1} & \multicolumn{1}{c|}{48.1} & \multicolumn{1}{c|}{48.4} & 57.5 & \multicolumn{1}{c|}{80.1*} & \multicolumn{1}{c|}{86.1*} & \multicolumn{1}{c|}{80.4*} & 130.8* \\
\multicolumn{1}{c|}{3} & Ictal + Interictal & \multicolumn{1}{c|}{104.2} & \multicolumn{1}{c|}{104.1} & \multicolumn{1}{c|}{104.6} & 874.8 & \multicolumn{1}{c|}{138.6} & \multicolumn{1}{c|}{138.2} & \multicolumn{1}{c|}{138.8} & Undetected \\
\multicolumn{1}{c|}{4} & Artifact & \multicolumn{1}{c|}{-} & \multicolumn{1}{c|}{-} & \multicolumn{1}{c|}{-} & - & \multicolumn{1}{c|}{-} & \multicolumn{1}{c|}{-} & \multicolumn{1}{c|}{-} & - \\
\multicolumn{1}{c|}{5} & Ictal + Interictal & \multicolumn{1}{c|}{203.1} & \multicolumn{1}{c|}{228.3} & \multicolumn{1}{c|}{203.8} & 202.9 & \multicolumn{1}{c|}{147.8} & \multicolumn{1}{c|}{147.5} & \multicolumn{1}{c|}{148.2} & 155.6 \\
\multicolumn{1}{c|}{6} & Ictal & \multicolumn{1}{c|}{178.9} & \multicolumn{1}{c|}{201.9} & \multicolumn{1}{c|}{179.3} & 178.6 & \multicolumn{1}{c|}{-} & \multicolumn{1}{c|}{-} & \multicolumn{1}{c|}{-} & - \\
\multicolumn{1}{c|}{7} & Ictal & \multicolumn{1}{c|}{44.1} & \multicolumn{1}{c|}{43.8} & \multicolumn{1}{c|}{44.3} & 55.9 & \multicolumn{1}{c|}{-} & \multicolumn{1}{c|}{-} & \multicolumn{1}{c|}{-} & - \\
\multicolumn{1}{c|}{8} & Ictal & \multicolumn{1}{c|}{43.9} & \multicolumn{1}{c|}{43.9} & \multicolumn{1}{c|}{44.3} & 62.8 & \multicolumn{1}{c|}{-} & \multicolumn{1}{c|}{-} & \multicolumn{1}{c|}{-} & - \\
\multicolumn{1}{c|}{9} & Ictal & \multicolumn{1}{c|}{41.7} & \multicolumn{1}{c|}{26.1} & \multicolumn{1}{c|}{49.7} & 38.4 & \multicolumn{1}{c|}{-} & \multicolumn{1}{c|}{-} & \multicolumn{1}{c|}{-} & - \\
\multicolumn{1}{c|}{10} & Ictal & \multicolumn{1}{c|}{74.5} & \multicolumn{1}{c|}{74} & \multicolumn{1}{c|}{74.7} & 98.7 & \multicolumn{1}{c|}{-} & \multicolumn{1}{c|}{-} & \multicolumn{1}{c|}{-} & - \\ \hline
\multicolumn{2}{c|}{\begin{tabular}[c]{@{}c@{}}\textbf{Mean ± SD} \\\textbf{ per channel}\end{tabular}} & \multicolumn{1}{c|}{85.5 ± 60.3} & \multicolumn{1}{c|}{89.0 ± 71.2} & \multicolumn{1}{c|}{86.7 ± 59.8} & 180.3 ± 251.7 & \multicolumn{1}{c|}{122.2 ± 30.0} & \multicolumn{1}{c|}{123.9 ± 27.0} & \multicolumn{1}{c|}{122.5 ± 30.0} & 143.2 ± 12.4 \\ \hline
\multicolumn{2}{c|}{\textbf{Mean ± SD}} & \multicolumn{4}{c|}{110.4 ± 143.3} & \multicolumn{4}{c}{126.6 ± 27.9}
\end{tabular}}
\end{table}
\section{Discussion}

The paper provided empirical results corroborating our initial claim that neuromorphic devices can be a suitable processing unit for neural implants. Overall, NET-TEN proved to be an effective tool to detect seizure-like activity in LFP signals pre-recorded \textit{in vitro}, adding to the list of neuromorphic platforms or full custom small-scale systems previously employed to process electrophysiological signals for pattern recognition and anomaly detection. Note that the developed device is a special-purpose hardware whose design followed the specifications dictated by the particular targeted application, i.e., implantability, so minimization of area and power was prioritized. Indeed, area and power consumption of the final fabricated network were in line with the requirements of implantable technologies. As a reference, the average power consumed by the device during one detection task was comparable to the one required by the metabolic activity of a single biological neuron cell, which is estimated to consume about 0.5-4.0 nW \cite{kiyatkin2004brain}. Such ultra-low power performance stems from mainly three factors. First of all, the spike-based nature of computation makes dynamic power the predominant component in power consumption, that is drawn only in correspondence of a spike at the input, while static power dissipation due to the transistors leakage current lies in the order of pW and is thus negligible in comparison. Secondly, and to complement the first point, thanks to the relative low-bandwidth of LFP data, all the information content is adequately represented by sparse spike trains where the closest pulses are separated by a distance of 350 $\mu$s (to better understand, look at how LFPs are converted in Section \ref{sec:encoding}). Once again, the fewer the spikes, the lower the power consumption will be due to less internal switching activities of the transistors. Finally, the inherent error resilience of spiking neural networks enables the use of sub-threshold design methodologies and, as a result, voltage supply can be drastically reduced. \par

Another guiding principle in the design of NET-TEN was biological resemblance. Even today, it is still debated which features of the neuron and synapse grant neural circuits the ability to recognize patterns, or at what scale collective computational properties emerge. The solution in this regard was to opt for circuits that mimicked their biological analogues as faithfully as possible, and also to favor high versatility, allowing the behavior of single functional blocks to be modified at a later stage. Said flexibility encompassed the firing patterns of the neuron, the magnitude of depression/potentiation, the temporal dependence of the synaptic update and the strength of the coupling between delivered input spikes and neurons of the first layer. Looking at the individual components, the neuron acts as a high pass filter, since a spike is triggered only when the charge accumulated on $C_{v}$ is enough to cause the transmembrane potential to pass the switching threshold of the comparator, and this condition occurs exclusively when multiple incoming spikes follow one after the other. Therefore, the neuron output firing frequency is always smaller than the frequency of its weighted and accumulated pre-synaptic spike trains. Meanwhile, the EPSC circuit functions as an exponentially decaying kernel which implements the temporal convolution of the pre-synaptic spikes. For what concerns the STDP, given the very small capacitance value chosen to store the synaptic weight ($C_{w}$), the module ultimately works more like a dynamically evolving filter than a learning algorithm with a stable outcome. It follows that, in this case, it is improper to speak of a training phase and a testing phase; or rather, these terms simply do not apply here, because the weight cannot be retained for extended periods of time and thus presenting the network with a sample is not going to affect how the network reacts to successive samples. However, the circuit can be expected to react in a similar way for similar input patterns.  \par 

A key determinant of the success of the whole system resides in how the analog signals were translated into spikes before being fed to NET-TEN. Generally speaking, an encoding algorithm must preserve all the task-relevant information and ideally do it with the least possible amount of spikes. This paper applied the SFE algorithm, which is a temporal spike encoding scheme that makes use of a threshold. A previous study conducted on a smaller network that shared the same basic units (neuron, EPSC and STDP) as NET-TEN demonstrated the feasibility of detecting ictal events regardless of the chosen threshold value and it also disclosed a trade-off between threshold and power consumption \cite{ronchini2021cmos}. However, things change if the dataset incorporates a greater degree of biological variability, as in this paper. When samples present different baseline noise, as evident from the comparison between Figure \ref{fig:I_II_A}a and \ref{fig:I_II_A}b, a fixed threshold results in couples (UP and DW) of input spike trains with heterogeneous firing rates. At this point, identifying a unique set of biases that can lead to the same detection accuracy for all samples becomes challenging. Similarly, the threshold can be optimized for each single sample by choosing the one that minimizes the error between original and reconstructed signal, but in the absence of epileptiform activity  (e.g., when we want to observe an artifact) the resulting threshold tends to be very low and prompt an excessive number of spikes. Ultimately, these experimental observations suggest that, for a predetermined set of biases, NET-TEN behaves optimally if the encoded samples all have approximately equal spike density. Under these circumstances, the device can be said robust to biological variability. 	

NET-TEN identified the pathological activity with low-latency, achieving a much lower detection delay than previously proposed neural-prosthetic seizure detectors \cite{salam2011novel,chen2013fully,altaf20151}. In its current state, the device is not yet able to differentiate between ictal and interictal events and unfortunately is not robust to artifacts. \textit{Per contra}, the paper pinpointed two elements that would help address these issues. Ictal and interictal discharges could be told apart by the number of spikes emitted at the output, which is consistently higher for ictal events compared to interictal. Instead, electrical artifacts could be isolated on the basis of the ISI between the first output spikes. A bump circuit would certainly be instrumental in both situations to compare the firing rate of the output neurons to a target value \cite{delbrueck1993bump}. Another approach that could be adopted to accomplish artifact rejection is the network-level mechanism presented by Burelo et al. \cite{burelo2021spiking} who introduced a global-inhibitory neuron in the architecture to constantly suppress the activity of the output layer neurons, together with a dis-inhibitory neuron that gets excited by the input UP and DW spikes and silents the global-inhibitory neuron. The entire mechanism relies on the fact that the excitation of the disinhibitory neuron due to an artifact is too brief to culminate in the repression of the global-inhibitory neuron firing, so in the presence of an artifact, the activity of the output neurons keeps being inhibited. This method might not be ideal here since, how emerges from Figure \ref{fig:Boxplot}b, the interictal discharges are associated with an emission of even less spikes with respect to the artifact and would then be suppressed as well. Anyway, the dimension of the dataset is too small to draw any firm conclusions, and one single artifact cannot be sufficient to determine the statistics of its class.\par

Fully-analog subthreshold design comes with advantages and disadvantages. On one hand, working in the subthreshold regime amplifies the impact of process variations on the electrical response of the circuit. In conventional devices, this effect is unwanted and needs to be minimized through various strategies, but in neuromorphic systems, random device mismatch is desirable. The mismatch naturally generates a normal distribution of the network parameters and increases the probability that neurons will be attuned to specific input patterns, a phenomenon already known in scientific literature \cite{richter2015device,thakur2016low,sharifshazileh2021electronic}. Outside a given range the mismatch can have a detrimental nature and lead to unpredictable behavior, for instance it can give rise to neurons always spiking. This is the reason why only four out of ten output channels were considered to be informative for pattern recognition. Keep in mind that what has been said until now refers to the situation in which a fixed set of nominal biases is used throughout the measurements and also that, in the particular case of NET-TEN, there is no way of knowing \textit{a priori} which output neurons will prove useful to complete the task. On the other hand, subthreshold design hinders the drivability of the circuit. Such effect could be compensated by an increase in the transistors sizes, thereby compromising on the area, or otherwise by accepting the low fan-out and embracing a sparsely-connected architecture, as has been done here. \par 

The findings of this investigation represent a proof of concept that neuromorphic processors can identify electrophysiological biomarkers on-chip, in real-time and following an unsupervised learning paradigm. Thanks to its low-area, low-power and low-latency detection, the developed device is well suited for clinical and therapeutic applications, although further studies are needed to demonstrate the effectiveness of NET-TEN in a closed-loop sensing and conditioning system and to examine the dynamics arising from the interplay with the epileptogenic neural tissue. Finally, neuromorphic computing could play a central role in bringing to life the next generation of smart embedded sensors, that would perform data-driven and event-based computation at the edge, and this study is but one example.

\section{Methods}

\subsection{Brain slice preparation and maintenance}

Horizontal hippocampus-cortex (CTX) slices, 400~µm thick, were prepared from male CD1 mice 4-8 weeks old. Epileptiform activity was induced by treatment of the K+ channel blocker 4AP (250 µM). A detailed description of the methods can be found in  \cite{Panuccio2018}. All procedures have been approved by the Institutional Animal Welfare Body and by the Italian Ministry of Health (authorization 860/2015-PR), in accordance with the National Legislation (D.Lgs. 26/2014) and the European Directive 2010/63/EU. All efforts were made to minimize the number of animals used and their suffering.


\subsection{Multi-electrode array recording}

Extracellular field potentials were acquired using the Mc\_Rack software through a 6 x 10 planar MEA (Ti-iR electrodes, diameter 30 µm, inter-electrode distance 500 µm, impedance $<$ 100kΩ) connected to a MEA1060 amplifier (all from Multichannel Systems, Germany). The brain slices were continuously perfused at $\sim$ 1ml/min with artificial cerebrospinal fluid containing 4AP (\textit{cf.} \cite{Panuccio2018} for composition), equilibrated with carbogen, and maintained at 32° C. The signals were sampled at 2 kHz and low-passed at half the sampling frequency before digitization.

\subsection{Data Encoding}
\label{sec:encoding}

Discrete-time LFP data were encoded in MATLAB using the SFE algorithm defined in \cite{petro2019selection}. For each 100s long LFP sample, two arrays of length 100s*2kHz were declared and their elements were initialized to '0'. One array was allocated to store UP spikes and the other for DW spikes. The algorithm was initialized by taking the first value of the LFP signal as baseline. At every new time step, the signal was compared to the baseline $\pm$ a threshold. When the value exceeded the upper limit of the interval an UP spike was registered by assigning a '1' to the corresponding element of the corresponding vector and the baseline was updated to the baseline plus the threshold. Likewise, a DW spike was recorded if the lower limit was surpassed and the baseline was shifted to baseline minus threshold. The threshold value determines the sensitivity of the encoder to changes in the signal and the resulting spiking frequency, with smaller values leading to higher spike densities. Here it was selected following an iterative process as the one that returned an output with an average firing rate of approximately $30\%$, where the output was the logic sum of UP and DW sequences. A second script in Python converted the two digital streams into real waveforms. Each '1' was replaced by a pulse of 130 $\mu s$ full width at half maximum and 200 mV amplitude and the rest was zero padded. Finally, said waveforms were resampled at 10kHz, saved as two separate columns in a text file and uploaded on a Keysight 33500A3 arbitrary waveform generator.

\subsection{Neuron and synapse models}

Silicon neurons were based on the mathematical model of spiking neuron developed by Izhikevich \cite{izhikevich2003simple}. The model is derived from the bifurcation analysis of the Hodgkin–Huxley model and retains its most salient features, thereby enabling the faithful reproduction of a wide range of firing patterns commonly observed in biological cortical neurons. Izhikevich neurons are described by the following two-dimensional system of ordinary differential equations:

\begin{subequations}
\begin{align}[left = \empheqlbrace\,]
& \dfrac{dV_{mem}}{dt}=0.04{V_{mem}}^{2}+5V_{mem}+140-U+{I}_{syn} \label{eqn:IzA} \\[10pt]
& \dfrac{dU}{dt}=a(bV_{mem}-U) \label{eqn:IzB}
\end{align}
\end{subequations}

With the after-spike reset defined as

\begin{equation}
\label{eqn:Iz_reset}
\text{if}\;\;\; V_{mem} \geq 30 mV:
\begin{cases}
& V_{mem} \text{ is set to } c \\
& U \text{ is incremented by } d
\end{cases}
\end{equation}

Where $V_{mem}$ is the transmembrane voltage, that is the difference in electric potential between the interior and the exterior of the cell membrane, and $U$ is a state variable named the slow membrane recovery variable. Unlike the Hodgkin–Huxley neuron, the model does not explicitly characterize the electrodiffusion kinetics of single specific ionic channels. However, the slow variable $U$ reflects the overall state of activation of the potassium channels and inactivation of the sodium channels \cite{izhikevich2003simple}.
According to \ref{eqn:IzA}, $U$ acts as a negative feedback for $V_{mem}$. In Equation \ref{eqn:IzB}, $a$ can be regarded as a time constant which determines the rate of evolution of $U$, and $b$ is another parameter that defines the strength of the coupling between $V_{mem}$ and $U$, meaning how much the variation of the transmembrane voltage influences the dynamics of the slow variable. Instead, $c$ and $d$ from Equation \ref{eqn:Iz_reset} are the after-spike reset values for the two state variables \cite{izhikevich2003simple}. \par
Synapses adhered to the STDP rule, according to which synaptic modification depends on the temporal relationship between pre- and postsynaptic firing. This dependency holds true only if the distance between pre- and postsynaptic spike is within an interval of tens of milliseconds. According to this algorithm, when a pre-synaptic spike anticipates a post-synaptic spike the synaptic efficacy is strengthened; on the contrary, when the pre-synaptic neuron fires after the post-synaptic neuron, the connection between them is weakened \cite{song2000competitive}. In mathematical terms, STDP is modeled as:

\begin{equation}
W(\Delta t) = \begin{cases}
A_+ e^{-\frac{\Delta t}{\tau_+}} &\text{if $\Delta t \geq 0$}\\
-A_- e^{\frac{\Delta t}{\tau_-}} &\text{if $\Delta t < 0$}
\end{cases}
\label{eq:STDP}
\end{equation}

$W(\Delta t)$ is the synaptic modification; $\Delta t$ is the time elapsed from when the pre-synaptic neuron emits a spike to when the post-synaptic neuron spikes; $\tau_+$ and $\tau_-$ are time constants in the order of $10ms$, and the parameters $A_+$ and $A_-$ are hyper-parameters that set the synaptic weight change rate.

\subsection{Hardware design}

All CMOS circuits were designed and simulated in sub-threshold analog domain by means of Cadence$^{\circledR}$ Virtuoso$^{\circledR}$ platform. Full custom layout design of the chip was carried out using Virtuoso$^{\circledR}$ Layout Suite. Calibre was used to perform the Design Rule Check and verify whether the geometric constraints imposed by the semiconductor manufacturer were met. The same tool was employed for the Layout Versus Schematic verification and to extract the parasitic resistance and capacitance. Parasitics were then harnessed in post-layout simulations to ensure that each instantiated physical cell and its corresponding schematic circuit were functionally equivalent. The chip was fabricated by Taiwan Semiconductor Manufacturing Company in 180 nm process technology. The silicon die was packaged with a 48-pin Quad Flat No-Lead (QFN48) package. The ten output channels of the neuromorphic device were buffered on the PCB to increase their drivability. For this purpose, we chose the MAX44280 operational amplifier by virtue of its 0.4 pF input capacitance that limited the load at the output nodes. To minimize the parasitic capacitance introduced by the PCB traces, in the PCB layout, the buffers were placed as close as possible to the neuromorphic chip footprint. Circuit biases were controlled through adjustable voltage dividers on the PCB obtained from 100 kOhm potentiometers with 25 turns to allow fine adjustments. The maximum achievable value for each voltage bias was 200 mV, with one turn corresponding to a variation of 8 mV. $V_{DD}$ and $V_{D}$ could be tuned up to 450 mV, at the expense of the tuning precision, that dropped to 18 mV per turn. 

\subsection{Performance metrics}

To find the precision of the system, each measurement of NET-TEN output firing was compared with the label signal associated with the delivered input sample. Precision is defined as:

\begin{equation}
Precision = {\frac{TP}{TP+FP}}
\label{eq:Precision}
\end{equation}

Where TP means true positives and it is the number of spikes emitted by NET-TEN in correspondence of any pathological activity, that is when the label was either 1 (interictal) or 2 (ictal); instead, FP are the false positives, the number of spikes registered at the output when the input displayed healthy unaltered activity (baseline, label = 0). A precision score of 1.0 means that every retrieved event is clinically-relevant. The algorithm for computing the precision score looped over the whole dataset. Hence the score took into account the spike emitted for all samples and by all four significant output channels. 
The ictal detection delay was calculated as the time difference between the rising edge of the ictal event in the label signal and the rising edge of the first subsequent spike impulse in the recorded output. Specifically, the rising edge was the time stamp at which the signal crossed half of its maximum value. Same approach was adopted to compute the interictal detection delay.  
Considering the first two spikes appearing at the output of the neuromorphic device after the onset of an ictal episode, the interspike interval was estimated as the time elapsed between the falling edge of the first impulse and rising edge of the second one. Procedure used also for each of the interictal discharges.

\section{Funding statement}
This work was funded by the European Union through the projects: H2020-MSCA-IF-2014 RebUs - Rewiring Brain Units, GA 660689, awarded to GP. H2020-FETPROACT-2018(RIA) HERMES - Hybrid Enhanced Regenerative Medicine Systems, GA 824164.

\bibliographystyle{unsrt}
\bibliography{main}

\end{document}